\documentclass[showpacs, showkeys, pra,tightenlines,secnumarabic, 10pt]{revtex4}
\usepackage[english]{babel}
\usepackage{amsmath,amsthm}
\usepackage{amsfonts}
\usepackage{graphicx}
\usepackage{epsfig}

\theoremstyle{definition}

\theoremstyle{remark}

\numberwithin{equation}{section}

\newcommand{\eps}{\varepsilon}

\newcommand{\qubit}[1]{\ensuremath{{|#1\rangle}}}

\newcommand{\qpsi}{\ensuremath{{\qubit{\psi}}}}
\newcommand{\qphi}{\ensuremath{{\qubit{\varphi}}}}
\newcommand{\enum}[2]{\ensuremath{{#1_1#2#1_2#2\ldots#2#1_n}}}

\newcommand{\Ent}{\ensuremath{{E_{Hmin}}}}

\newcommand{\shannon}{\ensuremath{H_{sh}}}

\newcommand{\hmes}{\ensuremath{H_{meas}}}

\begin{document}

\title[Calculation of quantum discord and entanglement measures using the random mutations optimization algorithm]{Calculation of quantum discord and entanglement measures using the random mutations optimization algorithm}%
\author{A.Yu.\,\,Chernyavskiy}%
\email{andrey.chernyavskiy@gmail.com}
\affiliation{Institute of Physics and Technology, Russian Academy of Sciences, 117218, Moscow, Russia}
\affiliation{Faculty of Computational Mathematics and Cybernetics, Lomonosov Moscow State University, 119991, Moscow, Russia}



\begin{abstract}
The simple and universal global optimization method based on simplified multipopulation genetic algorithm is presented. The method is applied to quantum information problems. It is compared to the genetic algorithm on standard test functions, and also tested on the calculation of quantum discord and minimal entanglement entropy, which is an entanglement measure for pure multipartite states.
\end{abstract}
\maketitle
\section{Introduction}

Research in quantum information science often leads to global optimization problems, and one of the main examples of this is the calculation of numerical measures of quantum correlations. One class of such measures is entanglement measures of pure multipartite quantum states. For instance, the calculation of the geometric entanglement measures for pure states~\cite{wei2003geometric} and the calculation of the minimal measurement entropy \cite{chernyavskiy2009entanglement}, which is considered in this paper, represents the global optimization problems. The second one is measures of mixed states. Most of such measures are based on minimization. Sometimes it's possible to use special optimization methods like convex optimization~\cite{audenaert2001asymptotic}, but in most cases we need to use general algorithms. For example, global minimization is needed to calculate the relative entropy of entanglement~\cite{henderson2000information}, geometric measures of entanglement~\cite{wei2003geometric}, any convex roof of measure for pure states (for example, the entanglement of formation~\cite{wootters1998entanglement}).

The third type of quantum correlation measures is quantum discord, which was proposed independently by~\cite{ollivier2001quantum} and~\cite{henderson2001classical}. Quantum discord is based not on entanglement, but on the existence of non-commutative quantum measurements, so separable (not entangled) states can have non-zero discord. There is a supposition that not only entanglement, but also quantum discord can be the resource for computations that outperform classical ones~\cite{datta2008quantum}. The definition of quantum discord is based on the minimization, and the analytical calculations was made only for a few classes of quantum states~\cite{luo2008quantum,ali2010quantum,rulli2011global}, that's why it's perspective to use modern global optimization algorithms.

It's well known that there is no universal method for finding global extremum of an arbitrary function. Till last years gradient methods (Newton's method, gradient descent, conjugate gradient method, etc.) were the most popular for optimization problems~\cite{nocedal1999no}. However, these methods are adapted for finding a local extremum or a global extremum of very specific functions.

Nevertheless, recently the increasing computational power allows to use more resource-intensive, but more effective algorithms. Generally, such algorithms use biological or physical heuristics. The most popular algorithms are ant colony
optimization~\cite{dorigo1992ola}, genetic algorithm~\cite{holland1975ana}, particle swarm optimization~\cite{kennedy2001si}, quantum annealing~\cite{apolloni1989qso} and simulated annealing~\cite{kirkpatrick1984osa}.

We presents random mutations algorithm, which is the simplified genetic algorithm. The algorithm is being tested on standard global optimization problems and on the computation of the minimal measurement entropy, which is the entanglement monotone for pure multipartite states~\cite{chernyavskiy2009entanglement,myphd}. After that the algorithm is used to compute quantum discord and tested on two-qubit states with maximally mixed subsystems.

\section{Random Mutations Optimization}

\subsection{General algorithm}

Let's describe the random mutations minimization algorithm.

Consider a general real global optimization problem. For a given $n$-parameter function $f(\mathbf{x})$ (fitness function) we need to find a vector $\mathbf{x}=\{x_1,x_2,\ldots,x_n\}$ such that
\begin{equation}
f(\mathbf{x})=\min\limits_{y}f(\mathbf{y}).
\end{equation}

The general random mutations algorithm consists of the following steps:
\begin{enumerate}
  \item \label{initialization} \textbf{Initialization.} Generate $n_{pop}$ random real vectors $x_i$ of dimension $n,$ $i \in \{\overline{1,n_{pop}}\}$.
  \item \label{mutation} \textbf{Mutation}. From each $x_i$ generate a set $D_i$ of $n_{des}$ vectors of dimension $n$ (descendants of $x_i$). Each vector of $D_i$ is independently generated from $x_i$ by a probabilistic mutation process. We can also include $x_i$ into $D_i.$
  \item \label{selection} \textbf{Selection.} From each set $D_i$ we choose one ``winner'' vector and take it as $x_i$ for the next generation. Also we store the best one of new $x_i$ vectors (the vector with minimal $f$) as $x_{best}.$
  \item \label{termination} \textbf{Termination.} If the termination conditions are satisfied, we take $x_{best}$ as the result, else go to the step 2.
\end{enumerate}

As it can be noticed, this algorithm is very similar to multipopulation genetic algorithm without migration and crossover.

The most important and specific steps are mutation (\ref{mutation}) and termination conditions (\ref{termination}). Let's describe all steps.

\subsection{Algorithm steps}
\subsubsection{Initialization}
The process of the generation of initial vectors is probabilistic. We can choose different distributions according to a minimization problem. But almost always we can take uniform distribution on $[v_{min},v_{max}]$ for each component of $x_i.$

\subsubsection{Mutation}
We use some special adaptive type of mutation. First of all, we must choose the distribution $NM$ for the number of components being mutated. For example we can choose uniform distribution on $[1, n_{maxmut}].$ This is very important parameter of algorithm and, of course, must be chosen according to the length of the vector.

Every new vector $x_{ij} \in G_i$ is generated from $x_{i}$ by the following procedure: we randomly choose $n_{mut} \sim NM$ components of the vector $x_{i}$ and then independently change each of them by the rule
\begin{equation}
v_{new} = v_{old} + m \cdot b^{p},
\end{equation}
where $m$ (magnitude) and $p$ (power) are uniformly distributed on $[-1,1]$ and $[p_{min},p_{max}]$ respectively.  The $p_{max}$ parameter always depends on the minimization problem and is important. We usually take $p_{min} = -9$ and the base $b=10.$

Such type of mutation is very similar to decimal or binary coding in genetic algorithms. It helps to adapt mutation amplitude during the algorithm: in the beginning even big mutations can win, but when the algorithm is near minimum almost only small steps can decrease the target function. But it's important that even during the final steps algorithm can find an appropriate large mutation (this is good for challenging local minima problem).

\subsubsection{Selection}
We use the simplest selection process: the vector with minimal fitness is taken as a winner. Another selection can be used, for example, winner can be chosen probabilistically according to its fitness.

\subsubsection{Termination conditions}
Every standard termination condition can be used, but the best among experiments was the next one: the algorithm stops if for every of $n_{stall}$ last generations the $x_{best}$ was changed on the value lesser than $eps.$

This termination condition is very effective against local minimum problem.

\subsection{Features and parameters}
The main advantages of random mutations algorithm are simplicity and easy parallelization. As a consequence the algorithm can be effectively realized and used on high-performance distributed computing systems. As it will be shown further, in spite of its simplicity, random mutations can outperform more complex algorithms.

The main parameter influencing the work of the algorithm is $n_{pop}.$ Obviously, increase of $n_{pop}$ leads to higher convergence speed, but also increases the probability to converge to local extrema. The second important parameter is the number of descendants $n_{des}.$ Like population size parameter of genetic algorithms it strongly depends on the optimization problem and in most cases it must be tuned manually to obtain better results. The parameters $n_{maxmut}$ and $p_{max}$ can be chosen accordingly to the number of arguments and the scale of a fitness function.

Notice that $n_{pop}$ subpopulations of random mutations algorithm are not totaly independent, because they are linked by termination conditions. That's why the algorithm is not similar to the simple repeating of the single population optimization procedure many times. Some subpopulations can converge to ``bad'' local minimums, near which the convergence is slow, and termination conditions stops the algorithm in such cases.

\section{Testing on standard functions}
\label{part:StandartFunctions}

Random mutations algorithm was tested on different standard test functions~\cite{panait2002comparative, janson2005hierarchical}:

\begin{equation}
\begin{array}{c}
  F_{Rosenbrock}(x) = \sum\limits_{i=1}^{n-1} \left(100(x_{i+1}-x_i^2)^2 + (1-x_i)^2\right), \\
  \\
  \min\limits_{-2.048\leq x_i < 2.048} F_{Rosenbrock}(x) = F(1,1, \ldots,1) = 0;
\end{array}
\end{equation}

\begin{equation}
\begin{array}{c}
F_{Rastrigin}(x) = 10n + \sum\limits_{i=1}^{n} \left(x_i^2 - 10 \cos(2\pi x_i)\right),\\
\\
\min\limits_{-5.12\leq x_i < 5.12} F_{Rastrigin}(x) = F(0,0, \ldots,0) = 0;\\
\end{array}
\end{equation}

\begin{equation}
\begin{array}{c}
F_{Griewank}(x) = 1 + \sum\limits_{i=1}^{n} \frac{x_i^2}{4000}-\prod\limits_{i=1}^{n}\left(\cos(\frac{x_i}{\sqrt{i}})\right),\\
\\
\min\limits_{-512\leq x_i < 512} F_{Griewank}(x) = F(0,0, \ldots,0) = 0;\\
\end{array}
\end{equation}

\begin{equation}
\begin{array}{c}
F_{Schwefel}(x) = 418.98288727\cdot n - \sum\limits_{i=1}^{n} x_i \sin\left(\sqrt{|x_i|}\right),\\
\\
\min\limits_{-512\leq x_i < 512} F_{Schwefel}(x) = F(420.968750, \ldots,420.968750) = 0.\\
\end{array}
\end{equation}

\begin{figure}
\centering
\begin{tabular}{cc}
\epsfig{file=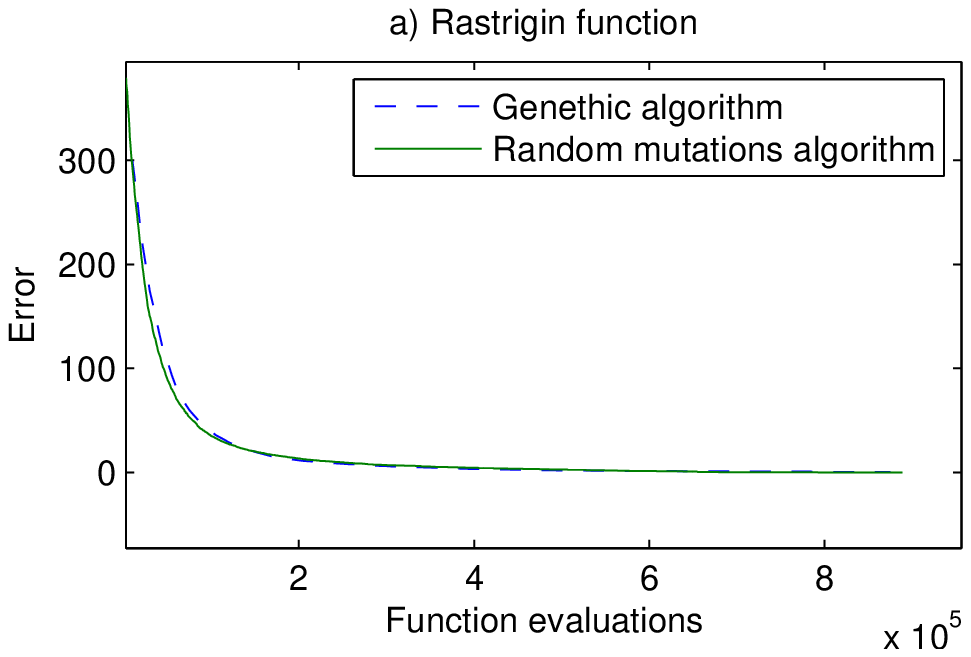,width=0.5\linewidth,clip=}&
\epsfig{file=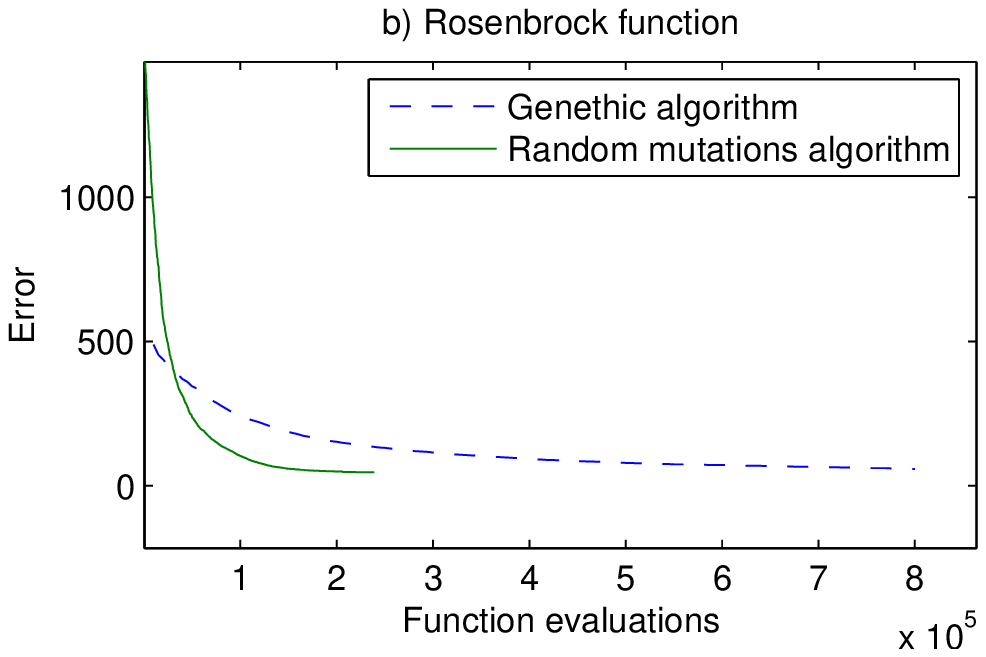,width=0.5\linewidth,clip=} \\
\bigskip\\
\epsfig{file=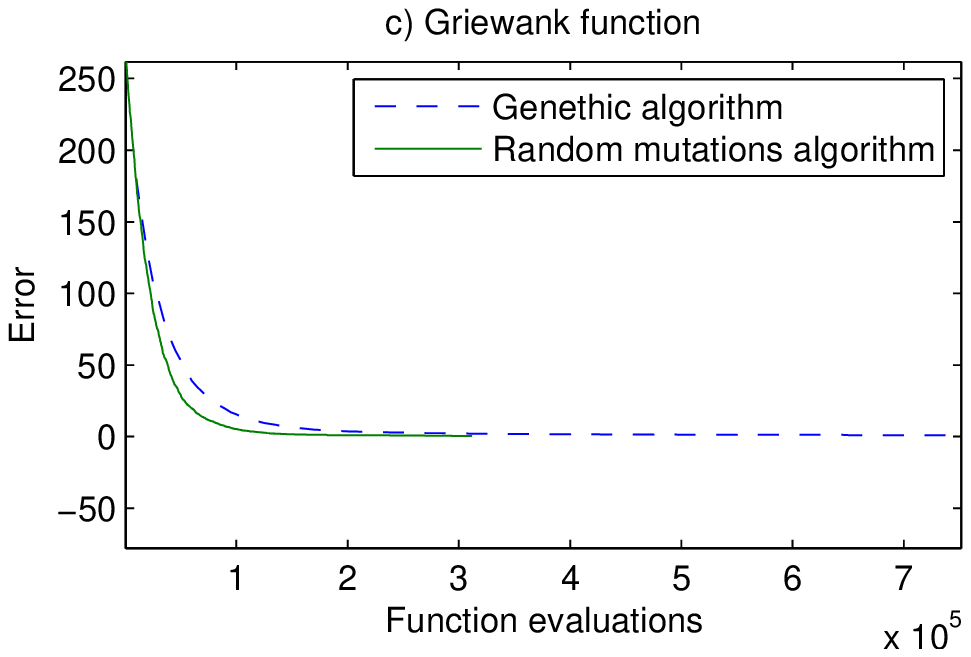,width=0.5\linewidth,clip=} &
\epsfig{file=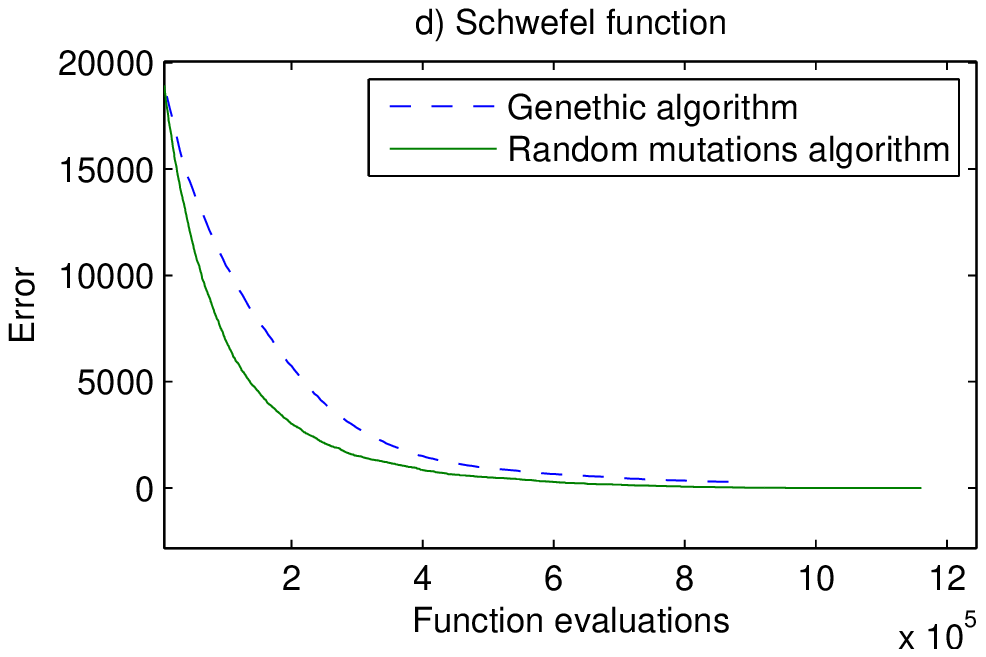,width=0.5\linewidth,clip=}
\end{tabular}
\caption{The convergence of the random mutations and genetic algorithm on standard test problems.}\label{fig:StandardTests}
\end{figure}

Of course, the assessment of the dependence of the accuracy from the number of algorithm iterations is not good, because the single iteration can be very difficult. So, we need to research the dependency of the accuracy from the number of fitness function evaluations. The main task of standard tests was only the verification of the correctness and competitiveness of the random mutations algorithm. The test functions was taken with the number of parameters $n=50.$ We used the standard MATLAB genetic algorithm. The size of the population was $5000,$ other parameters was standard. The parameters for random mutations were as follows: $n_{pop}=40,\: n_{maxmut}=5,$ for the Rosenbrock and Schwefel functions $n_{des}=20,$ for the Rosenbrock and Griewank functions $n_{des} = 10.$ For both algorithms $n_{stall} = 50, \: \eps = 10^{-6},$ but algorithms have different termination conditions. The MATLAB genetic algorithm stops if the averaged change of the best fitness over last $n_{stall}$ generations is lesser than $eps.$ The test results averaged over 20 experiments are plotted on Fig.~\ref{fig:StandardTests}. The graphs for each algorithm are presented for the minimal number of steps over experiments. We take $F_{Griewank}(400 \cdot x)$ to preserve scale The constraints of Schwefel function (as opposed to other test functions) is significant. Moreover the global minimum is closer to the bound than to the center of the search space, so the scaling begin to influence results. We take $F_{Schwefel}(350 \cdot x)$ for genetic algorithm and $F_{Schwefel}(100 \cdot x)$ for random mutations. The best and average results over experiments are presented on Tab.~\ref{tab:StandardTests}.

\begin{table}[ht]
  \begin{center}
        \begin{tabular}{|l|c|c|c|c|}
          \hline
            & \textit{Rastrigin} & \textit{Rosenbrok} & \textit{Griewank} & \textit{Schwefel} \\
            \hline
          \textbf{Genetic algorithm}&
          $\begin{array}{c}f_{best} = 0.0781 \\ f_{avg}=0.1983 \end{array}$  &
          $\begin{array}{c}f_{best} = 37.7922 \\ f_{avg}=49.8762 \end{array}$ &
          $\begin{array}{c}f_{best} = 0.3377 \\ f_{avg}=1.0207 \end{array}$ &
          $\begin{array}{c}f_{best} = 18.4127 \\ f_{avg}=53.9818 \end{array}$ \\
          \hline
          \textbf{Random mutations}&
          $\begin{array}{c}f_{best} = 0.00002 \\ f_{avg}=0.00006 \end{array}$  &
          $\begin{array}{c}f_{best} = 32.7622 \\ f_{avg}=42.4590 \end{array}$ &
          $\begin{array}{c}f_{best} = 0.0194 \\ f_{avg}=0.1171 \end{array}$ &
          $\begin{array}{c}f_{best} = 0.0006 \\ f_{avg}=0.0007 \end{array}$ \\
          \hline
        \end{tabular}
    \caption{The best and average results of genetic and random mutations algorithm on test functions over 20 experiments.}
    \label{tab:StandardTests}
  \end{center}
\end{table}

\begin{figure}[h]
  \bigskip
  \includegraphics[width=300pt]{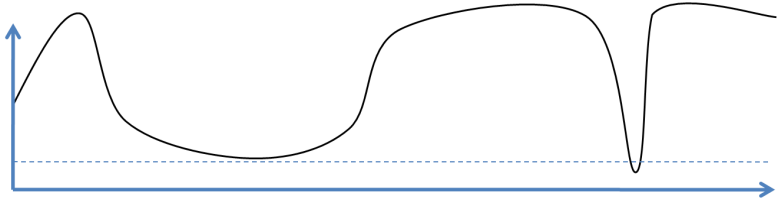}\\
  \caption{Function which is difficult for finding the global minimum.}\label{fig:HardFunction}
\end{figure}

This method of assessment is good for the test functions, which have regular structure. However, it is not so good for some real tasks. Consider a function with a local minima ``plateau'' and a global minima with a close value. The sample of such function is illustrated on Fig.~\ref{fig:HardFunction}.

Global optimization algorithms are probabilistic and with large probability converge to local minimums of such functions. But the probability of the correct answer is nonzero. So we need to assess the dependence of this probability from the number of fitness function evaluations. The exact method of such assessment will be described in Sec.~\ref{part:TestingHmin}.

\section{Calculation of minimal measurement entropy for pure states}

\subsection{Definition}

Let
\begin{equation}
\label{eq:PureState}
\qpsi=\sum\limits_{\enum{i}{,}=0}^{1} \lambda_{\enum{i}{}} \qubit{\enum{i}{}}
\end{equation}
be a pure quantum $n$-qubit state (the definition for qudits is the same).

The real positive number
\begin{equation}
\label{eq:MeasurementEntropy}
\hmes(\qpsi) = -\sum\limits_{\enum{i}{,}=0}^{1} |\enum{i}{}|^2 log |\enum{i}{}|^2
\end{equation}
is called a \textit{measurement entropy}.

Then the minimal measurement entropy over all orthogonal local measurement basis sets
\begin{equation}
\label{eq:MinimalEnropy}
\Ent(\qpsi) = \min\limits_{\enum{U}{,}} \hmes(\enum{U}{\otimes}\qpsi)
\end{equation}
is an entanglement measure (monotone) \cite{chernyavskiy2009entanglement}. Here $U_i$ are unitary matrices.

\subsection{Calculation}
For an arbitrary state \ref{eq:PureState} we need to find the global minima of $\hmes(\enum{U}{\otimes}\qpsi)$ over all possible sets of the $2\times 2$ unitary matrices $\enum{U}{,}.$

The parametrization of $2\times 2$ unitary matrices is well known:
\begin{equation}
U(\alpha, \beta,\delta,\gamma)=e^{i \alpha}\left(
              \begin{array}{cc}
                e^{i(-\beta-\delta)}\cos\gamma & -e^{i(-\beta+\delta)}\sin\gamma \\
                e^{i(\beta-\delta)}\sin\gamma & e^{i(\beta+\delta)}\cos\gamma \\
              \end{array}
            \right),
\end{equation}
where $\beta,\delta,\gamma$ are real numbers. We can easily see that $\alpha$ and $\beta$ doesn't affect $\hmes,$ so the final parametrization for our task is
\begin{equation}
U(\delta,\gamma)=\left(
              \begin{array}{cc}
                e^{-i\delta}\cos\gamma & -e^{i\delta}\sin\gamma \\
                e^{-i\delta}\sin\gamma & e^{i\delta}\cos\gamma \\
              \end{array}
            \right),
\end{equation}
and we need to solve the global minima problem for $2n$ real parameters.

If we want to calculate $\Ent$ for qudits ($d>2$) we can use the parametrization $U=e^{iH},$ where $H$ is Hermitian and can be obviously parameterized by $d^2$ real numbers.

\subsection{Testing and results}
\label{part:TestingHmin}
The main task of this work is testing of the random mutations algorithm, so we will present only test results, more calculations of $\Ent$ can be found in~\cite{chernyavskiy2009entanglement,myphd}.

The validity of calculations was tested on four cases:
\begin{enumerate}
  \item \textbf{Unentengled states.} Of course, for unentangled states minimal measurement entropy is equal to zero:
  \begin{equation}
      \Ent(\enum{U}{\otimes}\qubit{00\ldots0})=0.
  \end{equation}
  \item \textbf{GHZ states.} For the generalized GHZ-states $\qubit{GHZ}=\lambda_0\qubit{00\ldots0} + \lambda_1\qubit{00\ldots0}:$ \begin{equation}
      \Ent(\qubit{GHZ})=\shannon(\lambda_0,\lambda_1)=-|\lambda_0|^2 \log |\lambda_0|^2 - |\lambda_1|^2 \log |\lambda_1|^2.
      \end{equation}
  \item \textbf{Bipartite states.} For bipartite states $\Ent$ equals reduced von Neumann entropy.
  \item \textbf{Additivity.} For arbitrary $n$-qubits state $\qphi$ and $k$-qubits state $\qpsi$ for the $(n+k)$-qubits state $\qphi \otimes \qpsi:$
      \begin{equation}
      \Ent(\qphi \otimes \qpsi) = \Ent(\qphi)+\Ent(\qpsi).
      \end{equation}
\end{enumerate}

For all tests the random mutations algorithm gives correct results (tests were made for up to 17 qubits, the accuracy was $10^{-5}$). For these test states the algorithm shows a very good convergence. But some other states were very difficult, because the corresponding fitness function is of the type described in \ref{part:StandartFunctions} (with local minima ``plateau''). For example, for one state of 7-qubit Grover's algorithm the most possible result of the algorithm is $~4.0220,$ but there is another minimum $~3.968.$ We suppose that the second one is global, because of large amount of tests. Let's describe the method of quality assessment of global optimization algorithms on such problems, which was claimed in Sec. \ref{part:StandartFunctions}.

Consider $n_{exp}$ numerical experiments. It's obvious that the probability of the error decreases exponentially with repetition of optimization and we can take up the mean value of fitness calculations for achieving the success probability rate 0.5.

\begin{equation}
e_{0.5}=-n_{evaluations}/n_{exp} \cdot log_{0.5} (n_{err}/n_{exp}),
\end{equation}
where $n_{evaluations}$ is the summarized number of fitness function evaluations during all experiments, $n_{err}$ is the number of incorrect results. I.e. $n_{evaluations}/n_{exp}$ is the average number of fitness function evaluations during one experiment, $n_{err}/n_{exp}$ is the probability of error.

As there is no analytical values of $\Ent$ for computationally difficult states, we take the best result after $10000$ repeats of genetic, swarm particle, and random mutations algorithm for the ``correct'' global minimum.

Here is the example results for 17th step of 11-qubits Grover's algorithm.

\begin{itemize}
    \item genetic algorithm: $e_{0.5} \approx 5500$
    \item swarm particle optimization: $e_{0.5} \approx 11000$
    \item random mutations: $e_{0.5} \approx 3500.$
\end{itemize}

Other hard states give similar results.

Random mutations algorithm is very simple, but it demonstrates significant improvement for such problems. We must note that only a very simple type of swarm optimization was used, so different types and topologies probably may show better results.

Described tests show us that we can use random mutations algorithm to solve some difficult optimization problems of quantum information theory.

\section{Calculation of quantum discord}

\subsection{Definition}
Discord is a measure of non-classical correlations in quantum systems. Consider the bipartite mixed quantum state $\rho.$ Discord represents the difference between quantum mutual information and classical correlations:

\begin{equation}
D(\rho)=I(\rho)-C(\rho).
\end{equation}

The $I(\rho)=S(\rho_A)+S(\rho_B)-S(\rho)$ is the quantum mutual information. Here $S(\rho) = Tr \rho \log \rho$ is von Neumann entropy, $\rho_A$ and $\rho_b$ are reduced density matrices of subsystems.

The equation for classical correlations $C(\rho)$ is not so simple. First of all it's not symmetric. Let's choose subsystem $B$ and make some projective measurement with the observable $M,$ this measurement will destruct quantum correlations. Classical correlations after the measurement will be defined by $C_M(\rho) = S(\rho_A)-S(\rho|M),$ where $S(\rho|M)=\sum\limits_{i} p_i S(\rho_i),$ is a quantum conditional entropy, $p_i$ and $\rho_i$ are the probabilities and results of the measurement $M.$ To calculate $C(\rho)$ and discord itself we need to get the maximal $C_M(\rho),$ so $C(\rho) = S(\rho_A) - \min\limits_{M}S(\rho|M).$

\subsection{Testing}
We use random mutations optimization to calculate quantum discord. To do it we need to parameterize an observable $M.$
The simplest way is to parameterize $n \times n$ hermitian matrix by $n^2$ real numbers: $n$ on the diagonal and $n(n-1)$ above the diagonal. This method is redundant, because we only need the projectors of $M,$ but not its eigenvalues, but this redundancy only adds some additional work for the optimization algorithm, but it obviously doesn't affect the result.

The analytical results of Luo~\cite{luo2008quantum} were used to test the possibility of the random mutations algorithm to compute quantum discord of two-qubit systems . Consider a state

\begin{equation}
\label{eq:MaxMarginalsState}
\rho=\frac{1}{4}(I+\sum\limits_{j=1}^3 c_j \sigma_j \otimes \sigma_j).
\end{equation}

This is a state with maximally mixed subsystems. The quantum mutual information of this state is

\begin{equation}
I(\rho) = 2+\sum\limits_{i=0}^3 \lambda_i \log_2 \lambda_i,
\end{equation}
where
\begin{equation}
\begin{array}{c}
  \:\lambda_0=\frac{1}{4}(1-c_1-c_2-c_3), \:\: \lambda_1=\frac{1}{4}(1-c_1+c_2+c_3), \\
  \\
  \lambda_2=\frac{1}{4}(1+c_1-c_2+c_3), \:\: \lambda_3=\frac{1}{4}(1+c_1+c_2-c_3)
\end{array}
\end{equation}
are the eigenvalues of $\rho.$

The constraints of the coefficients $c_j$ are such that $0\leq \lambda_j \leq 1.$

For the classical correlations:
\begin{equation}
C(\rho)=\frac{1}{2}[(1-c)\log_2(1-c)+(1+c)\log_2(1+c)],
\end{equation}
where $c=max\{c_1,c_2,c_3\}.$

By definition, $D(\rho)=I(\rho)-C(\rho).$ It's important to notice that some important quantum spin systems can be described by \ref{eq:MaxMarginalsState}, so the value of discord for these systems can be calculated analytically~\cite{fel2012quantum}.

Let's describe tests carried out on such states. We simply generate random sets $\{c_1,c_2,c_3\},$ where each $c_j$ is uniformly distributed on $[-1,1]$. If the corresponding density matrix is valid we calculate discord for it and compare its value with the analytical result. More than 5000 states was checked. The parameters of algorithm was as follows: $n\_mut=5,\: n\_pop=10,\: n\_des=10.$ In all experiments the desired accuracy was achieved: difference between analytical and numerical results was lesser then $10^{-6}.$

\section{Conclusions}
The random mutations global optimization algorithm was presented, which is a kind of simplified multipopulation genethic algorithm. It gives the correct results and outperforms simple genetic optimization on standard tests and on the computation of minimal measurement entropy, which is the entanglement monotone for multipartite pure states. The algorithm was also used to compute quantum discord and tested on analytically known discord values of two-qubit states with maximally mixed subsystems. Thus, we can conclude that random mutations algorithm is perspective to solve some difficult problems of quantum information theory like calculation of discord and different entanglement measures for multipartite states.

\begin{acknowledgments}
The author is grateful to Prof.\,\,E.B.\,Feldman and Dr.\,\,S.I.\,Doronin for the initiation of research on quantum discord computation.

This work was partially supported by RFBR (project no. 12-01-31274).
\end{acknowledgments}

\bibliographystyle{unsrt}
\bibliography{bibliography}{}

\end{document}